\documentclass[12pt,preprint]{aastex6}

\newcommand{\lapprox} {\, \lower3pt\hbox{$\sim$}\llap{\raise2pt\hbox{$<$}}\,}
\newcommand{\gapprox} {\, \lower3pt\hbox{$\sim$}\llap{\raise2pt\hbox{$>$}}\,}

\begin{document}

\title{THE ROLE OF DIFFUSION IN THE TRANSPORT OF ENERGETIC ELECTRONS DURING SOLAR FLARES}

\author{Nicolas H. Bian\altaffilmark{1},
        A. Gordon Emslie\altaffilmark{2}, and
        Eduard P. Kontar\altaffilmark{1}}

\altaffiltext{1}{School of Physics \& Astronomy, University of Glasgow, Glasgow G12 8QQ, Scotland, UK (nicolas.bian@glasgow.gla.ac.uk)}

\altaffiltext{2}{Department of Physics \& Astronomy, Western Kentucky University, Bowling Green, KY 42101 (emslieg@wku.edu)}

\begin{abstract}

The transport of the energy contained in suprathermal electrons in solar flares plays a key role in our understanding of many aspects of flare physics, from the spatial distributions of hard X-ray emission and energy deposition in the ambient atmosphere to global energetics. Historically the transport of these particles has been largely treated through a deterministic approach, in which first-order secular energy loss to electrons in the ambient target is treated as the dominant effect, with second-order diffusive terms (in both energy and angle) being generally either treated as a small correction or even neglected.  We here critically analyze this approach, and we show that spatial diffusion through pitch-angle scattering necessarily plays a very significant role in the transport of electrons.  We further show that a satisfactory treatment of the diffusion process requires consideration of non-local effects, so that the electron flux depends not just on the local gradient of the electron distribution function but on the value of this gradient within an extended region encompassing a significant fraction of a mean free path.  Our analysis applies generally to pitch-angle scattering by a variety of mechanisms, from Coulomb collisions to turbulent scattering. We further show that the spatial transport of electrons along the magnetic field of a flaring loop can be modeled rather effectively as a Continuous Time Random Walk with velocity-dependent probability distribution functions of jump sizes and occurrences, both of which can be expressed in terms of the scattering mean free path.

\end{abstract}

\keywords{acceleration of particles -- Sun: activity -- Sun: flares -- Sun: X-rays, gamma rays}

\section{Introduction}

A solar flare involves a complex set of energy release and transport mechanisms, involving both non-thermal and thermal elements \citep[see, e.g.,][for reviews]{1988psf..book.....T,2011SSRv..159..107H,2011SSRv..159..301K}. A significant fraction \citep[e.g.,][]{2012ApJ...759...71E} of the energy released appears in the form of deka-keV electrons, the phase-space distribution of which varies with both energy and angle \citep[see, e.g.,][for a review]{2011SSRv..159..357Z}. The accelerated electrons lose energy principally through Coulomb collisions on ambient electrons \citep[e.g.,][]{1972SoPh...26..441B,1978ApJ...224..241E}, but additional processes associated with the turbulent environment through which they propagate \citep[e.g.,][]{2011A&A...535A..18B,2011ApJ...730L..22K} are also likely to be involved. Modeling of the Coulomb collision process has typically involved an analytic test-particle approach that principally involves systematic (secular) energy loss \citep[e.g.,][]{1971SoPh...18..489B,1972SoPh...26..441B,1978ApJ...224..241E}, although numerical solutions of the Fokker-Planck equation, involving collisional diffusion in pitch angle \citep[e.g.,][]{1981ApJ...251..781L,1991A&A...251..693M,1991ApJ...374..369B,2014ApJ...780..176K} and energy \citep[e.g.,][]{2014ApJ...787...86J} in addition to the secular energy loss term, have also been carried out.

The quality of the information contained in spatially resolved hard X-ray images from the {\em RHESSI} \citep{2002SoPh..210....3L} instrument has driven the need for a correspondingly greater fidelity and accuracy in describing the transport of the high-energy electrons responsible for the hard X-ray emission.  For example, \citet{2014ApJ...787...86J} have studied the variation of source size with energy through a Fokker-Planck analysis of the electron transport in a warm target, and \citet{2014ApJ...780..176K} have studied the influence of turbulent pitch-angle scattering on this source-size-with-energy relation. In such analyses, it has been generally accepted that at the energies necessary for hard X-ray emission, the electron dynamics is primarily controlled by secular (non-diffusive) energy loss.  On the other hand, the electrons responsible for {\it soft} X-ray emission are generally accepted to behave as part of a relaxed thermal distribution, with the predominant energy transport mechanism being spatial diffusion.

In this paper we critically assess this (as it turns out, unnecessarily dichotomous) paradigm by modeling, through a Fokker-Planck approach, the combined effects of collisional energy loss and angular diffusion of electrons in the hard X-ray energy domain.  In Section~\ref{fp-mfp} we introduce a transport equation that is valid at all electron energies and we show on the basis of this equation that it is {\it a priori} unjustified to adopt a purely deterministic view of transport of energetic electrons, even in a ``cold'' target. In Section~\ref{flux-local-gradient}, we discuss the relationship between electron flux and the gradient of the particle distribution function and we introduce the concept of non-local diffusion, in which, for sufficiently large mean free paths, the particle flux is determined not just by the local density gradient, but rather by a convolution of this gradient with a (velocity-dependent) spatial kernel, so that the problem becomes inherently non-local.  We then obtain an expression for this convolution kernel and show that it has a characteristic spatial extent that is a significant fraction of a mean free path. In Section~\ref{ctrw} we further show that the transport of energetic electrons along the guiding magnetic field can be viewed as a Continuous Time Random Walk with velocity-dependent probability distribution functions governing both the size and occurrence times of position-changing impulses. A summary and our conclusions are given in Section~\ref{conclusions}.

\section{The Fokker-Planck equation in a cold collisional target}\label{fp-mfp}

The one-dimensional kinetic equation for a gyrotropic ($\partial/\partial \phi = 0$) electron distribution function $f(z,\theta, v, t)$ (cm$^{-3}$ [cm~s$^{-1}$]$^{-1}$) along a guiding ambient magnetic field ${\bf B}$ is

\begin{equation}\label{kinetic}
\frac{\partial f}{\partial t} + v_\parallel \, \mathbf{b}.\nabla f = St^{v}(f) + St^{\theta}(f) + S \,\,\, ,
\end{equation}
where $z$ (cm) is the position of the particle gyrocenter along the magnetic field with direction ${\mathbf b} = {\mathbf B_0}/B_0$, $v_\parallel$ is the component of the electron velocity parallel to ${\mathbf b}$, and $S$ is the source of accelerated electrons injected into the transport region of interest, which in the flare context is called the target region. Transforming to the variables $(z,\mu,v,t)$, with $\mu= \cos \theta$ being the pitch-angle cosine, this Fokker-Planck equation may be rewritten as

\begin{equation}\label{fokker}
\frac{\partial f}{\partial t} + \mu v \, \frac{\partial f}{\partial z} = St^{v}(f) + St^{\mu}(f)+S(z,\mu,v) \,\,\, .
\end{equation}

The first term on the right side of Equation~(\ref{fokker}) is the operator associated with collisional scattering in velocity space:

\begin{equation}\label{stv}
St^{v}(f) = \frac{1}{v^{2}} \, \frac{\partial}{\partial v} \left [ v^{2} \left ( \nu_{C}(v) \, f v + D_{C}(v) \, \frac{\partial f}{\partial v} \right ) \right ] \,\,\, ;
\end{equation}
the two terms which appear on the right side describe, respectively, collisional friction and diffusion, with the velocity-space diffusion coefficient given by

\begin{equation}\label{diff-v}
D_{C}(v) = \frac{4 \pi n_{e}e^{4} \ln \Lambda \, k_B T_e}{m_e^3} \, \frac{1}{v^3} \,\,\, .
\end{equation}
Here $e$ (esu) and $m_e$ (g) are the electronic charge and mass, respectively, $n_e$ (cm$^{-3}$) is the density and $T_e$ (K) is the temperature, $k_B$ (erg~K$^{-1}$) is Boltzmann's constant, and $\ln \Lambda$ is the Coulomb logarithm.  The collisional friction coefficient $\nu_{C}(v)$ is related to the diffusion coefficient $D_{C}(v)$ by the ``fluctuation-dissipation'' relation

\begin{equation}\label{fd}
\nu_{C}(v) = \frac{4\pi n_e \, e^4 \, \ln \Lambda}{m_e^2} \, \frac{1}{v^3} \equiv \frac{m_e}{k_B T_e} \, D_C(v) \,\,\, ,
\end{equation}
a relation necessary for a Maxwellian distribution to satisfy the equilibrium condition $St^{v}(f)=0$.  This relation, together with inspection of the ratio of the terms on the right side of Equation~(\ref{stv}), implies that, in a non-equilibrium state and at energies $\frac{1}{2} m_e v^2 \gg k_B T_e$, the effect of deterministic friction generally dominates over diffusion.  Hence the transport of hard X-ray emitting electrons in flares has generally been largely treated through a deterministic approach, with only the first-order secular energy loss term used to describe the state of the system and diffusion terms in both angle and energy considered merely as small corrections.

While the above shows that in a cold target ($E \gg k_B T_e$) it is indeed justified to neglect the effect of {\it energy} diffusion in comparison with collisional friction, it turns out that {\it angular} diffusion is always of the same order of magnitude as (or, in the presence of turbulence, larger than) friction and hence cannot be neglected in any energy range.  To see this, we consider the second term on the right side of Equation~(\ref{fokker}), the pitch-angle scattering operator which strives to isotropize the distribution function.  This scattering operator takes the form

\begin{equation}\label{stmu}
St^{\mu}(f) =  \frac{\partial}{\partial \mu} \left [ D_{\mu\mu}\, \frac{\partial f}{\partial \mu} \right ] =\frac{\partial}{\partial \mu} \left [ \nu_C (v) \, (1-\mu^{2}) \, \frac{\partial f}{\partial \mu} \right ] \,\,\, ,
\end{equation}
and comparing this with the velocity-space diffusion term (Equations~(\ref{stv}) and~(\ref{fd})) shows that angular diffusion has an associated rate $\sim$$\nu_C$ and so occurs at a rate that is comparable with secular energy loss, even at non-thermal energies for which the energy diffusion term is negligible.  This means that it is {\it a priori} unjustified to adopt a purely deterministic description of transport of energetic electrons in a collisional cold target, a fact which considerably complicates the analysis\footnote{The solution to the standard deterministic cold target transport model can be obtained by e.g., straightforward application of the the method of characteristics \citep[see][and references therein]{2014ApJ...792....5E} because of the simple relation which exists between energy and time (or distance) in this case. No such simple relation is applicable in the presence of scattering.}.

From the collisional scattering frequency $\nu_C$, one can define the collisional mean free path

\begin{equation}\label{lambdac-def}
\lambda_{\rm C} (v) \equiv \frac{v}{\nu_{C}(v)} =\frac{m_e^2}{4\pi n_e \, e^4 \, \ln \Lambda} \,\, {v^4} \equiv \lambda_{ec} \left ( \frac{v}{v_{\rm te}} \right )^{4} \,\,\, ,
\end{equation}
where the thermal mean free path

\begin{equation}\label{lambda-ec}
\lambda_{\rm ec} = \frac{k_{B}^2 \, T_{e}^{2}}{\pi n_{e} e^4 \ln \Lambda}
\end{equation}
is the collisional mean free path at the thermal speed $v = v_{\rm te} \equiv \sqrt{2 k_B T_e/m_e} \,$. Although in the following we will mainly be interested in modeling the effect of collisional angular diffusion on spatial transport, the treatment is readily generalized to other mechanisms.  For example, the scattering frequency in Equation~(\ref{stmu}) could involve the sum of two components: collisional and turbulent \citep{2014ApJ...780..176K}, in which case we replace $\nu_C(v)$ by $\nu(v)=\nu_{C}(v)+\nu_{T}(v)$, so that the rate at which angular diffusion occurs is {\it greater} than the rate at which the electrons lose energy through Coulomb collisions.  Using the relation $\nu = v/\lambda$, the effective mean free path can be written in this situation as $1/\lambda(v) = 1/\lambda_{C}(v)+1/\lambda_{T}(v)$.

\section{Relationship between electron flux and gradient; local and non-local diffusion}\label{flux-local-gradient}

Given that the frequency associated with angular scattering is (at least) comparable to the collision frequency $\nu_C$, one cannot speak of a scatter-free or free-streaming behavior of energetic hard X-ray emitting electrons, even in a cold target. While this forces us to abandon the simple deterministic approach to spatial transport, we will, however, show below that it is possible to describe collisional transport of \emph{fast} electrons in term of a spatial diffusion, either local or non-local, provided the electron distribution is close to isotropic.

We first explore the relationship between the electron flux and the spatial gradient of the electron distribution function $f(z,\mu,v)$ throughout the volume in which transport occurs.  We will do so in a steady-state approximation\footnote{Such steady state is possible only for $v>v_{\rm te}$ in the presence of a source of electrons, with the full collisional operator conserving the number of particles.} appropriate to a steady injection of particles (time-dependent effects are briefly considered in the Appendix).  Thus, setting $\partial/\partial t = 0$, we obtain, for a general scattering frequency $\nu(v)$ appropriate to a mix of collisional and turbulent processes,

\begin{equation}\label{fokker3}
\mu v \, \frac{\partial f(v, \mu, z)}{\partial z} = St^{v}(f) + \frac{\partial}{\partial \mu} \left [ \nu (v)\, (1-\mu^{2}) \, \frac{\partial f(v, \mu, z)}{\partial \mu} \right ]+S(v,\mu, z)  \,\,\, .
\end{equation}
Integrating over $\mu$, we obtain

\begin{equation}\label{start}
\frac{\partial q(v,z)}{\partial z}= St^{v}[f_{0}(v,z)]+S_{0}(v,z) \,\,\, ,
\end{equation}
where

\begin{equation}\label{ff}
q(v,z)=v\int _{-1}^{+1} d\mu \, \mu \, f(v,\mu,z)
\end{equation}
(cm$^{-2}$~s$^{-1}$~[cm~s$^{-1}$]$^{-1}$) is the particle flux per unit velocity carried by electrons with speeds $v$,

\begin{equation}\label{iso}
f_0(v,z) = \int_{-1}^{+1} d\mu \, f(v, \mu, z)
\end{equation}
(cm$^{-3}$~[cm~s$^{-1}$]$^{-1}$) is the isotropic part of the distribution, i.e., the density of particles per unit velocity $v$, and $S_{0}(v,z) = \int_{-1}^{+1}d\mu \, S(v, \mu, z$) is the isotropic part of the source of particles. Notice that the definitions of both $f_0(v,z)$ and $q(v.z)$ involve projections onto the two first Legendre polynomials, $P_0=1$ and $P_1=\mu$, respectively. This, together with the fact that Legendre polynomials are eigenfunctions of the angular scattering operator $St^\mu$, strongly suggests a Legendre polynomial expansion of $f(v,\mu,z)$, which will indeed be used below.

In the standard diffusive approximation the particle flux is \emph{locally} proportional to the spatial gradient of the density of particles, i.e., to the spatial gradient of the isotropic component of the phase-space distribution function: $q_{d}(v,z) = - \kappa(v) \, \partial f_{0}(v,z)/\partial z$.  The extent to which such a diffusion approximation applies depends on the degree of anisotropy of the distribution function $f(v, \mu, z)$. As we shall see, in a steady state the particle flux $q(z,v)$ can in general be written as a spatial convolution with the electron density gradient:

\begin{equation}\label{fl-gr}
q(v,z) = - \lambda(v) \, v \int \limits_{-\infty}^{\infty} dz' \, K(v,z-z') \, \frac{\partial f_{0}}{\partial z'} \,\,\, ,
\end{equation}
where $K(v,z)$ is a velocity-dependent spatial kernel. This equation has an equivalent Fourier-space ($\partial/\partial z \rightarrow ik$) representation

\begin{equation}\label{lambda-def}
{\tilde q}(v,k)  = - ik \, \lambda(v) \, v \, {\tilde K}(v,k) \, \tilde{f}_{0}(v,k) \,\,\, ,
\end{equation}
and our goal reduces to providing an expression for the Fourier-space kernel ${\tilde K}(v,k)$.

Similar considerations arise in the propagation of solar energetic particles in the interplanetary medium \citep[where the focus has been mainly on the description
of time retardation effects through the so-called ``telegraph equation'';][]{1993ApJ...403..377G, 1994ApJ...427.1035K, 1994JGR....9919301S, 2000JPlPh..64..507Z, 2013A&A...554A..59L, 2015ApJ...808..157M}; see also the Appendix.
Here, however, we are instead primarily concerned with possible non-local {\it spatial} effects in determining the relation between the flux and the density gradient in a \emph{steady-state}.  Nevertheless, as we shall see, a method of analysis similar to that of \citet{1993ApJ...403..377G} still yields fruitful results, and we therefore decompose both the particle distribution function $f(v,\mu,z)$ and the source function $S(\mu)$ into a series of Legendre polynomials:

\begin{equation}\label{legendre-expansion-equation}
f(v,\mu,z) = \sum f_{n}(v,z) \, P_{n}(\mu) \,\,\, ; \quad S(v,\mu,z) = \sum S_{n}(v,z) \, P_{n}(\mu) \,\,\, .
\end{equation}
Substituting these expansions into the Fokker-Planck equation~(\ref{fokker3}) and then isolating each (orthogonal) Legendre polynomial component, one obtains the following recurrence relation, valid at each value of $v$ and $z$:

\begin{equation}\label{recurrence-relation}
v \left [ \frac{n}{2n-1}\frac{\partial f_{n-1}}{\partial z}+\frac{n+1}{2n+3}\frac{\partial f_{n+1}}{\partial z} \right ] = St^{v}[f_{n}] - \nu(v) \, n(n+1) \, f_{n} +S_{n} \,\,\, .
\end{equation}

First setting $n=0$ we obtain

\begin{equation}\label{zero-order}
\frac{1}{3} \, v \, \frac{\partial f_1}{\partial z} =St^{v}[f_0] + S_0 \,\,\, .
\end{equation}
Comparing this result with Equation~(\ref{start}) reveals, not surprisingly in view of the fact that the expression~(\ref{ff}) for the particle flux $q(v,z)$ involves a projection of $f(v,\mu,z)$ onto $P_{1}(\mu) \equiv \mu$, that the flux $q(v,z)$ is proportional to the first anisotropic component $f_1$, i.e., $q(v,z) = \frac{1}{3} \, v \, f_1(v,z)$. Our goal therefore now becomes to find an expression for $f_1$ in terms of $f_0$, thus obtaining a closed equation that describes the spatial transport $q(v,z)$ of the main isotropic component $f_0(v,z)$. We consider the case of isotropic injection, since observations \citep[e.g.,][]{2006ApJ...653L.149K} suggest that this is a reasonable assumption for electron acceleration in solar flares.  We therefore set all the anisotropic source terms $S_{n > 0} = 0$ and make the further approximation $St^{v}[f_{n>0}]=0$. Representing the spatial derivatives through Fourier components (i.e., writing $\partial/\partial z = ik$) and expressing the mean free path $\lambda$ in terms of the collision frequency through the relation $\lambda = v/\nu$, we obtain the recurrence relation for the quantities $\tilde{f}_n(v,k)$, valid at each value of $v$ and $k$,

\begin{equation}\label{recurrence}
\tilde{f}_n = -ik \lambda \left [ \frac{1}{(n+1)(2n-1)} \, \tilde{f}_{n-1} + \frac{1}{n(2n+3)} \, \tilde{f}_{n+1} \right ] \,\,\, .
\end{equation}
This is the fundamental equation that we will use to achieve the desired goal of expressing $\tilde{f}_1$ in terms of $\tilde{f}_0$.

First, we set $n=1$ in Equation~(\ref{recurrence}), giving

\begin{equation}\label{first-iteration}
\tilde{f}_1 = (-ik\lambda) \left ( \frac{1}{2 \cdot 1} \, \tilde{f}_0 \right ) + (-ik\lambda) \left ( \frac{1}{1 \cdot 5} \, \tilde{f}_2 \right ) \,\,\, .
\end{equation}
We then set $n=2$ in Equation~(\ref{recurrence}) to express $\tilde{f}_2$ in terms of $\tilde{f}_1$ and $\tilde{f}_3$, thus driving Equation~(\ref{first-iteration}) to the form

\begin{equation}\label{f2-step}
\tilde{f}_1 = \frac{-ik\lambda}{2} \, \tilde{f}_0 + \frac{(-ik\lambda)^2}{1 \cdot 5} \left ( \frac{1}{3 \cdot 3} \, \tilde{f}_1 + \frac{1}{2 \cdot 7} \, \tilde{f}_3 \right ) \,\,\, .
\end{equation}
Grouping the $\tilde{f}_1$ terms gives

\begin{equation}\label{next-approx-f1}
\tilde{f}_1 \left ( 1 - \frac{(-ik\lambda)^2}{1 \cdot 5 \cdot 3 \cdot 3} \right ) = \frac{-ik\lambda}{2} \, \tilde{f}_0 + \frac{(-ik\lambda)^2}{1 \cdot 5 \cdot 2 \cdot 7} \, \tilde{f}_3 \,\,\, .
\end{equation}
We next use $n=3$ in Equation~(\ref{recurrence}) to express $\tilde{f}_3$ in terms of $\tilde{f}_2$ and $\tilde{f}_4$, giving

\begin{equation}\label{next-next-step}
\tilde{f}_1 \left ( 1 - \frac{(-ik\lambda)^2}{1 \cdot 5 \cdot 3 \cdot 3} \right ) = \frac{-ik\lambda}{2} \, \tilde{f}_0 + \frac{(-ik\lambda)^3}{1 \cdot 5 \cdot 2 \cdot 7} \,
\left ( \frac{1}{4 \cdot 5} \, \tilde{f}_2 + \frac{1}{3 \cdot 9} \, \tilde{f}_4 \right ) \,\,\, .
\end{equation}
Then using $n=2$ and $n=4$ in Equation~(\ref{recurrence}) to substitute for the $\tilde{f}_2$ and $\tilde{f}_4$ terms, respectively, we obtain

\begin{eqnarray}\label{next-next-regroup}
\tilde{f}_1 \left ( 1 - \frac{(-ik\lambda)^2}{1 \cdot 5 \cdot 3 \cdot 3} \right ) & = & \frac{-ik\lambda}{2} \, \tilde{f}_0 \, + \cr
& + & \frac{(-ik\lambda)^4}{1 \cdot 5 \cdot 2 \cdot 7 \cdot 4 \cdot 5} \,
\left ( \frac{1}{3 \cdot 3} \, \tilde{f}_1 + \frac{1}{2 \cdot 7} \, \tilde{f}_3 \right ) + \cr
& + & \frac{(-ik\lambda)^4}{1 \cdot 5 \cdot 2 \cdot 7 \cdot 3\cdot 9} \,
\left ( \frac{1}{5 \cdot 7} \, \tilde{f}_3 + \frac{1}{4 \cdot 11} \, \tilde{f}_5 \right ) \,\,\, .
\end{eqnarray}
Again grouping the $\tilde{f}_1$ terms gives

\begin{eqnarray}\label{next-next-approx}
\tilde{f}_1  &\,&  \!\!\!\!\!\!\!\!\!\!\!\!\!\! \left ( 1 - \frac{(-ik\lambda)^2}{1 \cdot 5 \cdot 3 \cdot 3} - \frac{(-ik\lambda)^4}{1 \cdot 5 \cdot 2 \cdot 7 \cdot 4 \cdot 5 \cdot 3 \cdot 3} \right ) = \frac{-ik\lambda}{2} \, \tilde{f}_0 \, + \cr
& + & \left ( \frac{(-ik\lambda)^4}{1 \cdot 5 \cdot 2 \cdot 7 \cdot 4 \cdot 5 \cdot 2 \cdot 7} + \frac{(-ik\lambda)^4}{1 \cdot 5 \cdot 2 \cdot 7 \cdot 3\cdot 9 \cdot 5 \cdot 7} \right ) \, \tilde{f}_3 + \cr
& + & \frac{(-ik\lambda)^4}{1 \cdot 5 \cdot 2 \cdot 7 \cdot 3\cdot 9 \cdot 4 \cdot 11} \, \tilde{f}_5  \,\,\, .
\end{eqnarray}
We could continue in this way, substituting for $\tilde{f}_3$ in Equation~(\ref{recurrence}) to get terms in $\tilde{f}_2$ and $\tilde{f}_4$, then further substituting for $\tilde{f}_2$ in Equation~(\ref{recurrence}) to get an addition term in $\tilde{f}_1$ (and in $\tilde{f}_3$). The additional $\tilde{f}_1$ term (which will involve $(-ik\lambda)^6$) can then be grouped with the terms on the left side, and so on.  For now, we simply write the limiting form of this process as

\begin{equation}\label{truncated}
\tilde{f}_1  \left ( 1 - \frac{(-ik\lambda)^2}{1 \cdot 5 \cdot 3 \cdot 3} - \frac{(-ik\lambda)^4}{1 \cdot 5 \cdot 2 \cdot 7 \cdot 4 \cdot 5 \cdot 3 \cdot 3} + \ldots \right ) = \frac{-ik\lambda}{2} \, \tilde{f}_0 \,\,\, ,
\end{equation}
which leads to the following Fourier-space relation between the flux $\tilde{q}(v,k) \equiv (1/3) v \tilde{f}_1(v,k)$ and $\tilde{f}_{0}(v,k)$:

\begin{equation}\label{flux-expression-general}
{\tilde q}(v,k) = \frac{\frac{-ik\lambda}{6}}{ \left ( 1 - \frac{(-ik\lambda)^2}{1 \cdot 5 \cdot 3 \cdot 3} - \frac{(-ik\lambda)^4}{1 \cdot 5 \cdot 2 \cdot 7 \cdot 4 \cdot 5 \cdot 3 \cdot 3} + \ldots \right ) } \, v \, \tilde{f}_0(v,k)
\end{equation}
or equivalently to the following expression (see Equation~(\ref{lambda-def})) for the Fourier transform of the convolution kernel $K(v,z)$,

\begin{equation}\label{lambdak}
{\tilde K}(v,k)=\frac{1}{ 6 \left ( 1 - \frac{(-ik\lambda)^2}{1 \cdot 5 \cdot 3 \cdot 3} - \frac{(-ik\lambda)^4}{1 \cdot 5 \cdot 2 \cdot 7 \cdot 4 \cdot 5 \cdot 3 \cdot 3} + \ldots \right ) }\,\,\, .
\end{equation}
Keeping only the lead term in the denominator of (\ref{flux-expression-general}) results in the relation

\begin{equation}\label{diffusive-flux-fourier}
{\tilde q}(v,k) = \frac{-ik\lambda}{6} \, v \, \tilde{f}_0 (v,k) \,\,\, ,
\end{equation}
or the equivalent real-space expression

\begin{equation}\label{flux-local}
q(v,z) = -\frac{\lambda(v)}{6} \, v \, \frac{\partial f_{0}(v,z)}{\partial z} \,\,\, .
\end{equation}
This is the standard diffusion model, in which the particle flux carried by particles of speed $v$ is simply related to the local spatial gradient of the phase-space density of particles with that speed (i.e., Fick's law).

Generally, however, a non-local transport regime prevails.  To see this, let us now add the next term in the denominator of Equation~(\ref{flux-expression-general}), giving

\begin{equation}\label{flux-expression-non-local}
{\tilde q}(v,k) = \frac{\frac{-ik\lambda}{6}}{ \left ( 1 - \frac{(-ik\lambda)^2}{45} \right ) } \, v \, \tilde{f}_0(v,k) \,\,\, .
\end{equation}
The equivalent real-space differential equation relating $q$ and $f_0$ is of second order\footnote{adding subsequent terms to the general relation gives higher order differential equations}

\begin{equation}\label{quadratic}
q - \frac{\lambda^2}{45} \, \frac{\partial^2 q}{\partial z^2} = - \frac{\lambda v}{6} \, \frac{\partial f_{0}}{\partial z} \,\,\, .
\end{equation}
Noticing that Equation~(\ref{flux-expression-non-local}) is the product of a Lorentzian and the Fourier transform of the quantity $\partial f_{0}/\partial z$, we can take the inverse Fourier transform to obtain the convolution

\begin{equation}\label{del}
q(v,z) = - \frac{\sqrt{45}}{12} \int dz' \, \exp \left ( -\frac{\sqrt{45} \, |z-z'| }{\lambda(v)} \right ) \, v \, \frac{\partial f_{0}(v,z')}{\partial z'} \,\,\, .
\end{equation}
This expression generalizes the local expression~(\ref{flux-local}) to a non-local flux-gradient expression that involves a convolution of the gradient $\partial f_0/\partial z$ with a Laplacian (bi-exponential) distribution characterized by the $e$-folding scale length $\lambda(v)/\sqrt{45}$. At small mean free paths $\lambda(v)$ the Laplacian becomes very localized in the vicinity of $z = z^\prime$ and the particle flux reduces to the local expression~(\ref{flux-local}).

\section{Continuous time random walk of energetic electrons}\label{ctrw}

We have seen above that scattering, particularly when it is weak ($k \lambda \gg 1$), generally does not produces a spatial diffusion but rather (Equation~(\ref{del})) a non-local diffusion involving a convolution product with a spatial kernel $K(v,z)$ having a characteristic width of the order of the mean-free path $\lambda(v)$ (or, more accurately, $\lambda(v)/\sqrt{45}$). We now proceed to show that these results can also be obtained by modeling the electron dynamics as a velocity-dependent jump process in space, i.e., a Continuous Time Random Walk (CTRW), and that the standard diffusion limit is recovered when the distribution of jump sizes becomes very narrow.

Let us therefore consider the following Langevin equations for the dynamics of an electron: $\dot{z}=\zeta(t)$, $\dot{v} = - \nu_{C}(v) \, v$, where $\zeta(t)$ is an external noise source. When the latter is taken to be Gaussian, with zero mean $\langle \zeta(t) \rangle =0$ and white in time, $\langle \zeta(t)\zeta(t') \rangle = 2\kappa(v)\delta(t-t')$, this Langevin equation describes the dynamics of a strongly scattered electron undergoing a velocity-dependent spatial diffusion while systematically losing momentum to the background medium at a rate $\nu_C$.  Here we consider instead the following spatial dynamics, describing a series of finite spatial jumps, with the $k$th jump having a size $a_k$ (cm) and an occurrence time $t_k$:

\begin{equation}\label{stochastic-impulse-train}
\dot{z} = \zeta(t) = \sum_k \, a_k \, \delta(t-t_k) \,\,\, .
\end{equation}
The time intervals between neighboring pulses $\Delta t_{k}=t_{k}-t_{k-1}$ are assumed statistically independent and distributed according to an exponential probability distribution function (PDF)

\begin{equation}\label{time-distribution-poisson}
w_t(v,\Delta t)=\nu_a(v) \, e^{-\nu_a \Delta t} \,\,\, ,
\end{equation}
so that, equivalently, the occurrence times of the impulses are distributed within a Poisson distribution with mean $\nu_a^{-1}$.  The jump amplitudes $a_k$ are also considered to be statistically independent, with a PDF $w_a(v,a)$. This model thus considers $\zeta(t)$ as a shot noise process, which is white in time and has Poisson statistics, i.e., a Poisson white noise. The process $z(t)$ is also known as a continuous time random walk (CTRW) \citep{2000PhR...339....1M}; we note that a CTRW in {\it velocity} space has been proposed by \citet{2008ApJ...687L.111B} to describe the stochastic acceleration of electrons during solar flares.

We now consider the Chapman-Kolmogorov equation associated with this Markov process, focusing initially on the spatial part of the dynamics and thus (at this stage in the argument) ignoring velocity changes associated with the frictional term ${\dot v} = -\nu_C(v) \, v$. The fundamental quantity in this Chapman-Kolmogorov equation is the transition probability $W(z \, | \, z')$, which describes the probability per unit time that the particle jumps from position $z'$ to $z$. The evolution of the probability $P(v,z,t)$ that the particle of speed $v$ is at position $z$ at time $t$ is simply obtained by the sum $\int dz' \, W(z \, | \, z') \, P(v,z',t)$ of all transitions that bring particles of speed $v$ to $z$ from other positions $z^\prime$, minus the quantity $[ \, \int dz' \, W(z' \, | \, z) \, ] \, P(v,z,t)$, representing the sum of all transitions that cause electrons with speed $v$ to leave point $z$:

\begin{equation}\label{chapman-kolmogorov}
\frac{\partial P(v,z,t)}{\partial t} = \int dz' \, W(z \, | \, z') \, P(v,z',t) - \int dz' \, W(z' \, | \, z) \, P(v,z,t) \,\,\, .
\end{equation}

Now the transition probability $W(z \, | \, z')$ depends only on the difference between its arguments, $W(z \, | \, z') = W(z-z')$. Further, since we assume that the time intervals between jumps are drawn from a Poisson distribution characterized by the frequency $\nu_a(v)$, and that the jumps have an amplitude PDF $w_a(a)$, it follows that $W(z \, | \, z')=\nu_a(v) \, w_a(v,z-z')$ and the Chapman-Kolmogorov equation becomes

\begin{equation}\label{chapman-kolmogorov-next}
\frac{\partial P(v,z,t)}{\partial t} = \nu_a(v) \, \left [ \int dz' \, w_{a}(v,z') \, P(v,z-z',t) \, - \, P(v,z,t) \right ] \,\,\, .
\end{equation}
This is a non-local diffusion equation involving the convolution product of the probability density $P(v,z,t)$ with a kernel $w_a$ that represents the probability distribution function of jump sizes $a$.

We now recall the transport equation~(\ref{fl-gr}) and use it to write the evolution of the isotropic part of the distribution function $f_0(z,v,t)$ using the continuity equation:

\begin{equation}\label{flux-continuity}
\frac{\partial f_0(v,z,t)}{\partial t} = - \frac{\partial q(v,z,t)}{\partial z} = \frac{\partial }{\partial z} \left [ \lambda(v) \, v \int dz' \, K(v,z-z') \, \frac{\partial f_0(v,z,t)}{\partial z'} \right ] \,\,\, .
\end{equation}
Since the probability distribution function $P(v,z,t)$ and the dominant (isotropic) part of the phase-space distribution function $f_0(v,z,t)$ are equal to within a multiplicative constant (the total number of particles in the system), we can write Equation~(\ref{flux-continuity}) as

\begin{equation}\label{probability-continuity}
\frac{\partial P(v,z,t)}{\partial t} = \frac{\partial }{\partial z} \left [ \lambda(v) \, v \int dz' \, K(v,z-z') \, \frac{\partial P(v,z,t)}{\partial z'} \right ] \,\,\, .
\end{equation}

Equating the right sides of Equations~(\ref{chapman-kolmogorov-next}) and~(\ref{probability-continuity}), and taking the Fourier transform of the result, we see that

\begin{equation}\label{important-result}
\nu_{a}(v) \left ( \tilde{w}_{a}(v,k)-1 \right ) = - \, k^{2} \, \lambda(v) \, v \, {\tilde K}(v,k) \,\,\,
\end{equation}
or

\begin{equation}\label{omega-k}
\tilde{w}_{a}(v,k) = 1-\frac{k^{2} \, \lambda(v) \, v}{\nu_a(v)} \,\, \tilde{K}(v,k) = 1-\frac{ [k \lambda(v)]^2}{\nu_a(v) \, \tau(v)} \,\, \tilde{K}(v,k) \,\,\, ,
\end{equation}
where we have set $\tau(v) = \lambda(v)/v$.

Since we wish the CTRW model to replicate the non-local diffusion results of Section~\ref{flux-local-gradient}, we substitute for $\tilde{K}(v,k)$ from Equation~(\ref{lambdak}), keeping terms up to order $k^2 \lambda^2$ in the denominator, viz.

\begin{equation}\label{lambdak-truncated}
{\tilde K}(v,k) = \frac{1}{ 6 \left ( 1 + \frac{ k^2\lambda^2}{45} \right ) } \,\,\, .
\end{equation}
Substituting in Equation~(\ref{omega-k}) gives

\begin{equation}\label{matching-approaches}
\tilde{w}_{a}(v,k) = \frac{1 + k^2 \lambda^2 \left ( \frac{1}{45} - \frac{1}{6 \, \nu_a(v) \, \tau(v)} \right )}{1 + \frac{k^2 \lambda^2}{45}} \,\,\, .
\end{equation}

While this equation can obviously be satisfied for a variety of combinations $(\nu_a, \tilde{w}_a)$, we now, for reasons that will become quickly apparent, make the identification

\begin{equation}\label{constraint}
\nu_a(v) = \frac{15}{2 \, \tau(v)} = \frac{15 \, v}{2 \, \lambda(v)} \,\,\, ,
\end{equation}
corresponding to the temporal probability distribution function (Equation~(\ref{time-distribution-poisson}))

\begin{equation}\label{time-distribution-poisson-specific-value}
w_t(v,\Delta t) = \frac{15 \, v}{2 \, \lambda(v)} \, \exp \left ( - \, \frac{15 \, v \, \Delta t}{2 \, \lambda(v)} \right ) \,\,\, .
\end{equation}
This choice of the mean rate $\nu_a(v)$ nullifies the coefficient of the $k^2 \lambda^2$ term in the numerator of Equation~(\ref{matching-approaches}), resulting in the relatively simple expression

\begin{equation}\label{omega-expression-fourier}
\tilde{w}_{a}(v,k) = \frac{1}{ 1 + \frac{ k^2 \lambda^2(v)}{45}}
\end{equation}
for the Fourier transform of the amplitude probability distribution function.  Such a Lorentzian form corresponds to a real-space amplitude distribution function that has the Laplacian form

\begin{equation}\label{omega-expression}
w_{a}(v,z)=\frac{\sqrt{45}}{\lambda(v)} \exp { \left ( - \frac{\sqrt{45} \, \vert z \vert }{\lambda(v)} \right ) } \,\,\, ,
\end{equation}
which has the desirable property (for a probability distribution function) of being straightforwardly normalizable.  This simple form of $w_a(v,z)$ is the justification for the choice of mean occurrence rate $\nu_a(v)$ in Equation~(\ref{constraint}).

Equations~(\ref{time-distribution-poisson-specific-value}) and~(\ref{omega-expression}) specify respective distributions of impulse times and amplitudes that correspond to the nonlocal diffusive analysis of Section~\ref{flux-local-gradient}.  The velocity impulses $a_k$ (Equation~(\ref{stochastic-impulse-train})) occur with a Poisson distribution of times $t_k$ with mean $\nu_a^{-1} = 2 \lambda(v)/15 v $, and with an amplitude distribution that takes the form of a Laplace distribution with characteristic width $\lambda(v)/\sqrt{45}$.

For completeness, we can add the deterministic (frictional) dynamics:

\begin{equation}\label{impulse-train-friction-drag}
\dot{z} = \sum_{k} \, a_{k} \, \delta(t-t_{k}) \, ; \qquad \dot{v} = - \nu_C (v) \, v \,\,\, ,
\end{equation}
so that the full Chapman-Kolmogorov equation, allowing particles to change velocity as a result of collisions, becomes

\begin{equation}\label{chapman-full}
\frac{\partial P(v,z,t)}{\partial t} = \nu_a \left [ \int dz' w_{a}(v,z') \, P(v,z-z',t) \, - P(v,z,t) \right ] - \frac{\partial \left [ \nu_C(v) \, v \, P(v,z,t) \right ]}{\partial v} \,\,\, .
\end{equation}

Yet another equivalent way of writing the dynamics of one electron is the integral form

\begin{equation}\label{introduce-time}
P(v,z,t)=\delta(z) \, W_{t}(v,t) + \int _{0}^{t} \int_{-\infty}^{+\infty} w_{t}(v,t-t') \, w_{a}(v,z-z') \, P(v,z',t') \, dz' \, dt' \,\,\, ,
\end{equation}
where

\begin{equation}\label{cumulative}
W_{t}(t) = \int_{t}^{\infty} dt' \, w_{t}(v,t') \,\,\, .
\end{equation}
In Fourier-Laplace space, with the usual Laplace transform $\tilde{f}(s)=\int_{0}^{\infty} dt \, e^{-st}f(t)$, the previous integral equation transforms into

\begin{equation}\label{fourier-laplace-result}
\tilde{P}(v,k,s) = \tilde{W_{t}}(v,s) + \tilde{w_{t}}(v,s) \, \tilde{w_{a}}(v,k) \, \tilde{P}(v,k,s) \,\,\, .
\end{equation}
This yields the Montroll-Weiss equation

\begin{equation}\label{montroll-weiss}
\tilde{P}(v,k,s) = \frac{\tilde{W_t}(v,s)}{1-\tilde{w_{t}}(s) \, \tilde{w_{a}}(k)} = \frac{1-\tilde{w_{t}}(v,s)}{s} \, \frac{1}{1-\tilde{w_{t}}(s) \, \tilde{w_{a}}(k)} \,\,\, ,
\end{equation}
where we have used the fundamental relation $\tilde{W_{t}}(v,s)=(1-w_{t}(v,s))/s$ relating the Laplace transforms of a function and its derivative. Expanding this equation, we obtain the series representation

\begin{equation}\label{series-rep}
\tilde{P}(v,k,s)=\tilde{W_{t}}(v,s) \, \sum_{n=0}^{\infty} \, \left [ \tilde{w_{t}}(v,s)\tilde{w_{a}}(v,k) \right ]^n,
\end{equation}
which in real space corresponds to the Green's function

\begin{equation}\label{green-function}
P(v,z,t)=\sum_{n=0}^{\infty} \, b_{n}(v,t) \, c_{n}(v,z) \,\,\, .
\end{equation}
Here the functions $b_{n}$ and $c_{n}$ involve n-fold convolutions of the waiting-time and jump probability distribution functions, i.e., $b_{n}=W_{t}*w_{t}^{*n}$ and $c_{n}=w_{a}^{*n}$. The $n$-fold convolution of the exponential distribution $w_{t}$ is the Gamma distribution, viz.

\begin{equation}\label{gamma-distribution}
w_{t}^{*n}=\frac{\nu_{a}^{n} \, t^{n-1}}{(n-1)!} \, e^{-\nu_{a}t} \,\,\, ,
\end{equation}
and so we have

\begin{equation}\label{time-dependent-expression}
P(v,z,t) = e^{-\nu_{a}(v)t} \, \sum_{n=0}^{\infty} \, \frac{[\nu_{a}(v)t]^n}{n!} \,\, c_{n}(v,z) \,\,\, ,
\end{equation}
where $c_{n}(v,z)$ is the $n$-fold convolution of the Laplacian distribution~(\ref{omega-expression}).

Since the electron dynamics in energy space is deterministic, we know $v(t)=v(v_{0},t)$ by applying the standard method of characteristics. Hence the above expression also gives the Green's function $P(v_{0},z,t)$ corresponding to the $z=0$ injection of an electron with $v=v_{0}$ into a scattering medium with energy loss.  And, since $w_{a}(v=0,z)=0$, we see that when the particle reaches $v=0$ its PDF stops evolving in time. This interesting property follows from the fact that the diffusion process is controlled (Equation~(\ref{omega-expression})) by the (velocity-dependent) mean free path $\lambda(v)\sim v^{4}$, which approaches zero as $v$ approaches zero.

Overall then, a physical picture of the transport of energetic electrons in a cold target during solar flares can be summarized as follows: an electron with speed $v_{0}$ is injected from a source at $z=0$ at time $t=0$; its PDF is initially a delta function in space. As time evolves the particle makes large jumps around the source location, in general crossing it many times, and its PDF spreads symmetrically in space (but generally \emph{not} in a Gaussian manner, a consequence of the velocity dependence of the mean free path which controls the PDF of the distribution of impulse amplitudes -- Equation~(\ref{omega-expression})). As time progresses, the particle loses energy to the target and hence its speed $v$ and collisional mean free path $\lambda(v)$ both decrease.  This causes the distribution function~(\ref{omega-expression}) for the impulse sizes to become more and more peaked around zero and the non-local terms in the expression~(\ref{del}) for the particle flux become negligible; particles now evolve according to a local diffusive behavior with small jumps.  Only then does the PDF tend to relax to a Gaussian in space. Eventually the electron loses all its energy to the target, the mean free path $\lambda(v)$ goes to zero and the non-local kernel function~(\ref{omega-expression}) becomes strongly peaked at zero: at this point the electron dynamics become frozen, leaving a final PDF $P(z;v_{0})$. The target is continuously replenished by new injected energetic electrons from the source, hence maintaining a quasi-steady state during the flare.

\section{Summary and Conclusions}\label{conclusions}

Historically the transport of hard-X-ray-emitting electrons during solar flares has been largely treated as a deterministic process. In this work we have shown that consideration of angular scattering means that a deterministic treatment of the transport of energetic electrons is {\it a priori} unjustified (even at quite low scattering rates).  A correct treatment requires a non-local, probabilistic view, in which the relation between the flux and the density gradient is non-local, involving a convolution product with a velocity-dependent kernel, the spatial extent of which is about one-sixth ($1/\sqrt{45}$) of the mean free path. We have also developed a model (Section~\ref{ctrw}) in which energetic electrons perform a Continuous Time Random Walk along the guiding magnetic field of the loop; this process is characterized by a velocity-dependent distribution of impulse sizes and, for suitable choices of the distributions of impulse sizes (Equation~(\ref{time-distribution-poisson-specific-value})) and arrival times (Equation~(\ref{omega-expression})), reproduces exactly the non-local diffusion results of Section~\ref{flux-local-gradient}.

The diffusive treatment of electron transport developed in this work is currently being used, in a separate work, to derive revised expressions for the spatial distribution of energy deposition rate by accelerated nonthermal electrons, as essential input to models of the thermal and hydrodynamic response \citep[e.g.,][]{2005ApJ...630..573A} of flaring loops to heating by fast electrons.  This modeling also incorporates, through revised expressions for the mean free path $\lambda$, the effects of reduction of heat transport by turbulent scattering \citep{2016ApJ...824...78B,2016ApJ...833...76B}. We believe these tasks are particularly timely, since expressions for the energy deposition rate available so far in the literature \citep[e.g.,][]{1973SoPh...31..143B,1978ApJ...224..241E,1980ApJ...235.1055E} typically consider scattering in a deterministic test-particle approach; as such they do not adequately take into account the inherent diffusional nature of the scattering process, even at the high electron energies appropriate to hard X-ray emission.

A diffusive picture of the transport of energetic electrons during flares also opens up quite a number of further issues. Particles performing a random walk tend to return frequently to their initial location which, in the context of flare-accelerated energetic electrons, is the acceleration site itself. This may be an important factor (as opposed to, e.g., magnetic mirroring) in determining why the spatial PDFs of the hard X-ray flux are often observed \citep[e.g.,][]{2008ApJ...673..576X,2008ApJ...673.1181K,2012A&A...543A..53G} to peak near the top of flaring loops.  Also, while in a deterministic view of transport an electron gains energy only once (at its formation in the acceleration region), an electron performing a random walk can return to the acceleration site many times. This considerably complicates the overall dynamics of hard-X-ray-emitting electrons, has important implications for the efficiency of the electron acceleration process \citep{2012ApJ...755...32G}, and may be a key element in addressing the long-standing ``number problem'' \citep[e.g.,][]{1971SoPh...18..489B,1976SoPh...50..153L} of the number of accelerated electrons required to produce a given hard X-ray burst.

\acknowledgments

AGE was supported by grant NNX10AT78G from NASA's Goddard Space Flight Center. NHB and EPK were supported by a STFC grant. We thank Gary Zank for useful comments.

\appendix

\section{Appendix -- Approach to the Steady-State Solution}
Retaining the time dependence in the electron kinetic equation, the recurrence relation (cf. Equation~(\ref{recurrence-relation})) between the Legendre components of order $n>1$ can be written in the following form

\begin{equation}\label{recurrence-time-dependent}
\left ( 1+\tau_{n} \, \frac{\partial }{\partial t} \right ) f_{n} = -\lambda \left [ \frac{1}{(n+1)(2n-1)}\frac{\partial f_{n-1}}{\partial z} \, + \frac{1}{n(2n+3)}\frac{\partial f_{n+1}}{\partial z} \right ] \,\,\, ,
\end{equation}
where the relaxation time-scale

\begin{equation}\label{tau-n}
\tau_n = \frac{1}{\nu \, n(n+1)} = \frac{1}{n(n+1)} \, \frac{\lambda(v)}{v} \,\,\, .
\end{equation}
Since $\tau_n$ is a strongly decreasing function of $n$, the highest order anisotropies are the fastest to decay toward zero and so we can legitimately consider an expansion about the lowest values of $n$ in determining the essential temporal evolution of the system. Introducing the operators

\begin{equation}\label{a-operators}
A_n = -\frac{\lambda \, \partial/\partial z} {1+\tau_n \, \partial/\partial t} \,\,\, ,
\end{equation}
we may rewrite the recurrence relation in the convenient form

\begin{equation}\label{recurrence-operator}
f_n = A_n \left [ \frac{1}{(n+1)(2n-1)} \, f_{n-1}+ \frac{1}{n(2n+3)} \, f_{n+1} \right ] \,\,\, .
\end{equation}
Writing the first few relations

\begin{equation}\label{recurrence1}
f_1 = A_1 \left [ \frac{1}{2 \cdot 1} \, f_0 + \frac{1}{1 \cdot 5} \, f_2 \right ] \,\,\, ,
\end{equation}

\begin{equation}\label{recurrence2}
f_2 = A_2 \left [ \frac{1}{3 \cdot 3} \, f_1 + \frac{1}{2 \cdot 7} \, f_3 \right ] \,\,\, ,
\end{equation}

\begin{equation}\label{recurrence3}
f_3 = A_3 \left [ \frac{1}{4 \cdot 5} \, f_2 + \frac{1}{3 \cdot 9} \, f_4 \right ] \,\,\, ,
\end{equation}
we may proceed as in the text and express the flux carrying component $f_1$ as

\begin{equation}\label{f1-operator-first-few-terms}
f_1 = A_1 \left [ \!\! \left [ \frac{1}{2 \cdot 1} \, f_0 + \frac{1}{1 \cdot 5} \, A_2 \left \{ \frac{1}{3 \cdot 3} \, f_1 +
\frac{1}{2 \cdot 7} \left [ A_3 \left ( \frac{1}{4 \cdot 5} \, f_2 + \frac{1}{3 \cdot 9} \, f_4 \right ) \right ] \, \right \} \, \right ] \!\! \right ] \,\,\, .
\end{equation}
Substituting for $f_2$ in terms of $f_1$ and $f_3$ using Equation~(\ref{recurrence2}) and grouping terms, we obtain

\begin{equation}
\left [ 1 - \frac{1}{1 \cdot 5 \cdot 3 \cdot 3} \, A_1 \, A_2 - \frac{1}{1 \cdot 5 \cdot 2 \cdot 7 \cdot 4 \cdot 5 \cdot 3 \cdot 3}  \, A_1 \, A_2^2 \, A_3 \right ] \, f_1 = \frac{1}{2} \, A_1 \, f_0 + \ldots \,\,\, ,
\end{equation}
which should be compared with Equation~(\ref{truncated}) in the main text. We again notice that the numerical coefficients in front of the higher order (pseudo)-differential operators decrease very rapidly, so that we may, to a high degree of accuracy, truncate this expression by taking $A_{3}=0$ (thereby effectively setting $f_3=0$) to obtain

\begin{equation}
\left [ 1 - \frac{1}{45} \, A_1 \, A_2 \right ] \, f_1 = \frac{1}{2} \, A_1 \, f_0 \,\,\, ,
\end{equation}
i.e.,

\begin{equation}\label{time-diffusion}
\left ( 1 + \tau_1 \, \frac{\partial}{\partial t} \right ) \,
\left [ 1 - \frac{1}{45} \,
\frac{\lambda^2 \, \partial^2/\partial z^2}
{ \left ( 1 + \tau_1 \, \partial/\partial t \right ) \,
\left ( 1 + \tau_2 \, \partial/\partial t \right ) } \right ] \, f_1 = - \frac{1}{2} \, \lambda \, \frac{\partial f_0}{\partial z} \,\,\, .
\end{equation}
The time-dependent "telegrapher" relation between the flux and the gradient of particles \citep{RevModPhys.61.41} is obtained by taking $A_{2}=0$ (thus effectively setting $f_{2}=0$), i.e.,

\begin{equation}\label{time-approach}
\left ( 1 + \tau_1 (v) \, \frac{\partial}{\partial t} \right ) \, f_1 (v,z) = - \frac{1}{2} \, \lambda(v) \, \frac{\partial f_0(v,z)}{\partial z} \,\,\, ,
\end{equation}
where explicit velocity dependencies of both the relaxation timescale $\tau_1$ and the mean free path $\lambda$ have been noted. The spatial dependence of the particle flux $f_1$ depends on the local spatial derivative of the isotropic component $f_0$, so that the flux is determined by local diffusion.  However, at large times $t \gg \tau_1 = 1/2 \nu(v) = \lambda(v)/2v$, the higher-order $f_3=0$ closure result~(\ref{time-diffusion}) yields the steady-state relation

\begin{equation}
\left [ 1 - \frac{\lambda^2(v)}{45} \, \frac{\partial ^{2}}{\partial z^{2}} \right ] \, f_1(v,z) = - \frac{1}{2} \, \lambda(v) \, \frac{\partial f_0(v,z)}{\partial z} \,\,\, ,
\end{equation}
in which the flux is now determined by non-local diffusion effects (cf. Equation~(\ref{flux-expression-non-local}) in the text).

\bibliographystyle{apj}
\bibliography{bian_et_al_diffusive_transport}

\end{document}